\documentclass[12pt]{elsarticle}

\usepackage[margin=1.0in]{geometry}
\usepackage{algorithm}
\usepackage{algorithmicx}
\usepackage[noend]{algpseudocode}
\usepackage{graphicx}
\usepackage{tikz}
\usetikzlibrary{automata, arrows.meta, positioning}
\usepackage{subcaption}
\usepackage{caption}
\usepackage{amsmath}

\usepackage{multirow}
\usepackage{array}
\usepackage{tabularx}
\usepackage{color,soul}
\newcolumntype{L}[1]{>{\raggedright\let\newline\\\arraybackslash\hspace{0pt}}m{#1}}
\newcolumntype{C}[1]{>{\centering\let\newline\\\arraybackslash\hspace{0pt}}m{#1}}
\newcolumntype{R}[1]{>{\raggedleft\let\newline\\\arraybackslash\hspace{0pt}}m{#1}}

\makeatletter
\def\ps@pprintTitle{%
    \let\@oddhead\@empty
    \let\@evenhead\@empty
    \def\@oddfoot{\footnotesize\itshape}%
    \let\@evenfoot\@oddfoot
    }
\makeatother

\begin{document}

\begin{frontmatter}

\title{Improving TAS Adaptability with a Variable Temperature Threshold}

\author{Anthony Dowling}
\ead{dowlinah@clarkson.edu}
\author{Ming-Cheng Cheng}
\author{Yu Liu}
\address{Department of Electrical and Computer Engineering, Clarkson University, 8 Clarkson Ave., Potsdam, 13699, New York, USA}

\begin{abstract}
   
    Thermal-Aware Scheduling (TAS) provides methods to
    manage the thermal dissipation of a computing chip
    during task execution. These methods aim to avoid issues
    such as accelerated aging of the device, premature
    failure and degraded chip performance. In this work, we
    implement a new TAS algorithm, VTF-TAS, which makes use
    of a variable temperature threshold to control task
    execution and thermal dissipation. To enable adequate
    execution of the tasks to reach their deadlines, this
    threshold is managed based on the theory of fluid
    scheduling~\cite{fluid_2019}. Using an evaluation
    methodology as described in~\cite{dowling_2023}, we
    evaluate VTF-TAS using a set of 4 benchmarks from the
    COMBS benchmark suite~\cite{combs_2020} to examine its
    ability to minimize chip temperature throughout schedule
    execution. Through our evaluation, we demonstrate that
    this new algorithm is able to adaptively manage the
    temperature threshold such that the peak temperature
    during schedule execution is lower than POD-TAS, with
    no requirement for an expensive search procedure to
    obtain an optimal threshold for scheduling.

\end{abstract}


\end{frontmatter}

\section{Introduction}
\label{sec:intro}

POD-TAS~\cite{dowling_2023}, is a TAS algorithm that relies
on a pair of static temperature thresholds. Through
evaluation, this algorithm proved to be very capable of
controlling the CPU temperature within the defined maximum
temperature. However, this algorithm carries a limitation in
its requirement for a fixed threshold. Requiring the
threshold to be static invokes a need for a search to obtain
an optimal value for the threshold. This search carries high
computational cost to find the optimal threshold values.
Instead, if the algorithm were able to find an optimal or
nearly optimal threshold during scheduling, this would avoid
the high cost of searching for the optimum. This could be
achieved with a number of granularities; the simplest would
be to make a more intelligent search capability for the
algorithm to perform scheduling as little as possible while
finding the optimum threshold for a given task set and CPU,
but this would still involve a great deal of cost. Moving to
finer granularities, the threshold could be optimized per
scheduling period, using the threshold found in the previous
period as an initial value as to a potential optimum when
starting a search. However, both of these methods would
require a search mechanism which would invoke high
computational cost.  Instead, a purely variable threshold
would be preferable.

In this work, we utilize concepts from
POD-TAS~\cite{dowling_2023}, including the TAS notation, to
define, implement, and test an algorithm that is capable of
dynamically optimizing the temperature threshold during
schedule creation.  To achieve dynamic adjustment of the
temperature threshold, a heuristic is developed based on
fluid scheduling~\cite{fluid_2019}.  Fluid scheduling is a
scheduling methodology that aims to maintain a constant rate
of execution for tasks so that execution completes exactly
at the task's deadline.  Our heuristic aims to maintain a
nearly fluid execution rate for all of the scheduled tasks,
maintaining their execution rate by adjustment of the
temperature threshold. If the tasks need to execute more to
maintain a fluid execution rate, the temperature threshold
is increased; but if the tasks are executing faster than
their fluid execution rate, the threshold is lowered.

\section{VTF-TAS: A Variably Thresholded Fluid TAS Algorithm}

The control flow of the VTF-TAS algorithm is similar to the
flow of POD-TAS.  However, the ASSIGN\_TASKS and
UPDATE\_STATES algorithm are both changed to meet the needs
of this new algorithm. Furthermore, the SELECT\_TASKS
algorithm is integrated into the new ASSIGN\_TASKS\_VTF
algorithm to allow for greater control over task to core
assignments.

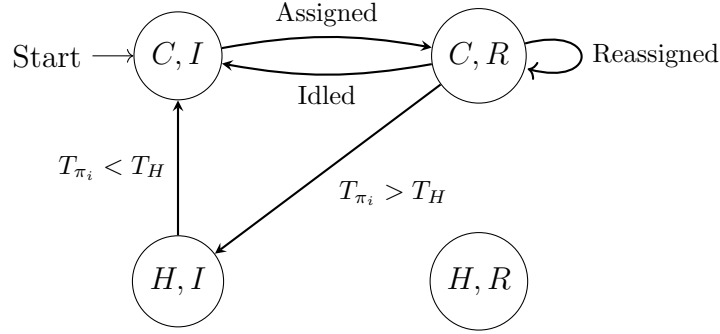
\begin{figure}
    \centering
    \begin{tikzpicture} [node distance = 2cm, on grid, auto]
        \node (ci) [state, initial, initial text = {Start}] {$C,I$};
        \node (cr) [state, right =4cm of ci] {$C,R$};
        \node (hi) [state, below =3cm of ci] {$H,I$};
        \node (hr) [state, below =3cm of cr] {$H,R$};

    \path [-stealth, thick]
        (ci) edge [bend left=10] node {\footnotesize Assigned} (cr)
        (cr) edge [bend left=10] node {\footnotesize Idled} (ci)
        (cr) edge [loop right] node {\footnotesize Reassigned} ()
        (cr) edge node {\footnotesize $T_{\pi_i}>T_H$} (hi)
        (hi) edge node {\footnotesize $T_{\pi_i}<T_H$} (ci)
        ;
    \end{tikzpicture}

    \caption{State Transition Diagram for CPU Cores in VTF-TAS}
    \label{ch:vtftas:fig:cpustates}
\end{figure}

\begin{algorithm}[t]
    \caption{Pseudocode of VTF-TAS Core State Management (Primary Concept)}
    \label{ch:vtftas:alg:states}
    \begin{algorithmic}[1]
        \Function{UPDATE\_STATES\_VTF}{$\pi$,$\Psi$,$T_{\pi}$,$\Lambda$}
            \For{$\pi_i\in\pi$}
                \If{$T_{\pi_i}<T_H$}
                    \State $run\_state\gets R$
                    \If{$\Lambda_{\pi_i} = \emptyset$}
                        \State $run\_state\gets I$
                    \EndIf
                    \State $\Psi_{\pi_i}\gets(C,run\_state)$
                \Else
                    \State $\Psi_{\pi_i}\gets(H,I)$
                \EndIf
            \EndFor
            \State \Return $\{\Psi_{\pi_i}\:\:\text{where}\:\:\pi_i\in\pi\}$
        \EndFunction
    \end{algorithmic}
\end{algorithm}

\begin{algorithm}[t]
    \caption{Pseudocode of the VTF-TAS Algorithm (Primary Concept)}
    \label{alg:vtftas}
    \begin{algorithmic}[1]
        \Function{VTF-TAS}{$\tau$, $T_{H0}$, $t_{end}$, $W_D$, $\pi$, $p_{len}$}
            \State $t_{curr}\gets0$;
            $\sigma\gets[]$;
            $T_H\gets T_{H0}$;
            $\Lambda\gets\emptyset$;
            $\Psi\gets\{(C,I) : \pi_i\in\pi\}$;
            \While{$t_{curr}<t_{end}$}
                \State $t_{prem}\gets p_{len}-(t_{curr}\%p_{len})$
                \If {$\text{NEED\_ASSIGNMENT}(\tau,\Psi,\Lambda,\pi,t_{prem})\lor t_{curr}=0\lor t_{prem}=0$}
                    \State $\Lambda\gets\text{ASSIGN\_TASKS\_VTF}(\tau, \pi, \Psi, t_{prem})$
                    \State $\text{\textit{append}}(\sigma, (t_{curr},\Lambda))$
                \EndIf
                \State $t_{curr}\gets t_{curr}+\Delta t$
                \State $T_H\gets\text{UPDATE\_THRESHOLD}(\tau,T_H,p_{len},t_{curr},W_D)$
                \State $T_{\pi}\gets\text{PREDICT\_TEMP}(P_{\sigma},t_{curr})$
                \State $\Psi\gets\text{UPDATE\_STATES\_VTF}(\pi,\Psi,T_{\pi},\Lambda)$
            \EndWhile
            \State \Return $\sigma$
        \EndFunction
    \end{algorithmic}
\end{algorithm}

To start, a key difference between VTF-TAS and
POD-TAS~\cite{dowling_2023} begins with its use of CPU core
states. In POD-TAS, there were three thermal states: cold,
warm, and hot. In VTF-TAS, we do not use the $T_C$
threshold; thus, we only have cold and hot for available
thermal states. Idle and running remain for execution
states, however, giving us 4 possible CPU core states. The
CPU core states used in VTF-TAS and their transitions are
shown in Figure~\ref{ch:vtftas:fig:cpustates}. Again, the
$(H,R)$ state is unreachable, because as soon as the
temperature of a core is detected to have exceeded $T_H$, it
is idled.  Algorithm~\ref{ch:vtftas:alg:states} demonstrates
this logic in the for of pseudocode to construct the set
$\Psi$, which is used in other algorithms to denote the CPU
core states.

\begin{algorithm}[t]
    \caption{Pseudocode of Checking for Assignment Required (Primary Concept)}
    \label{alg:needass}
    \begin{algorithmic}[1]
        \Function{NEED\_ASSIGNMENT}{$\tau$, $\Psi$, $\Lambda$, $\pi$, $t_{prem}$}
            \If{$\exists\tau_i\in\Lambda : TimeLeft(\tau_i)=0$}
            \Comment{Check if \textbf{any} \textit{assigned} tasks have finished}
                \State \Return true
            \EndIf
            \If{$\forall\tau_i\in\tau : TimeLeft(\tau_i)=0$}
            \Comment{Check if \textbf{all} tasks have finished}
                \State \Return false
            \EndIf
            \If{$\exists \pi_i\in\pi : (T_{\pi_i}>T_H\land\Lambda_{\pi_i} \neq\emptyset)$}
            \Comment{Check if \textbf{any} cores are overheated}
                \State \Return true
            \EndIf
            \If{$\exists\tau_i\in\tau : (TimeLeft(\tau_i)=t_{prem} \land \tau_i\notin\Lambda)$}
                \Comment{Check if \textbf{any} tasks need an}
                \State \Return true
                \Comment{override \underline{and} are not executing}
            \EndIf
            \If{$\exists\tau_i\in\tau : (TimeLeft(\tau_i)\neq0\land\tau_i\notin\Lambda)$}
                \Comment{Check if \textbf{any} tasks are incomplete}
                \If{$\exists\pi_i\in\pi : \Psi_{\pi_i}=(C,I)$}
                    \Comment{Check if \textbf{any} core state is $(C,I)$}
                    \State \Return true
                \EndIf
            \EndIf
        \EndFunction
    \end{algorithmic}
\end{algorithm}

The core VTF-TAS algorithm is shown in
Algorithm~\ref{alg:vtftas}.  Initially, the $T_H$ value is
given as an input to the algorithm as $T_{H0}$.  This allows
the algorithm to be ``hinted'' as to a valid initial value
of $T_H$.  This is also used to improve the algorithm
flexibility later on. The primary loop flows similarly to
POD-TAS~\cite{dowling_2023}, iterating through the time domain and constructing
the schedule, $\sigma$. However, VTF-TAS checks at each time
step if an assignment is needed using
Algorithm~\ref{alg:needass}. This checks the conditions of
the CPU and tasks to ensure that assignments are made when
required. After assignment, $T_H$ is updated depending on
the execution progress of the tasks, temperature prediction
is performed in the same manner as POD, and the CPU core
states are updated using
Algorithm~\ref{ch:vtftas:alg:states} before moving on to the
next time step. The key improvements of VTF-TAS arise from
the updated task assignment method and the ability for $T_H$
to be updated as the schedule is constructed.

\begin{algorithm}[t]
    \caption{Pseudocode of VTF-TAS Task to Core Mapping Algorithm (Primary Concept)}
    \label{ch:vtftas:alg:assign}
    \begin{algorithmic}[1]
        \Function{ASSIGN\_TASKS\_VTF}{$\tau$,$\pi$,$\Psi$,$t_{prem}$}
            \State $\pi_{sorted}\gets sort(\{\pi_i\in\pi| T_{\pi_i} \leq T_{\pi_{i+1}})$
            \Comment{Sort cores by ascending temperature}
            \State $\pi'\gets sort(\{\pi_i\in\pi| \Psi_{\pi_i}\neq(H,I) \land T_{\pi_i} \leq T_{\pi_{i+1}})$
            \Comment{Exclude overheated cores}
            \State $\tau_{sorted}\gets sort(\tau|TimeLeft(\tau_i)\geq TimeLeft(\tau_{i+1}))$
            \Comment{Sort tasks by remaining time}
            \State $\tau_\Omega\gets \{\tau_i\in\tau_{sorted}|TimeLeft(\tau_i)=t_{prem}\}$
            \Comment{Find overridden tasks}
            \State $\tau'\gets\tau_{sorted}-\tau_\Omega$
            \Comment{Find \textbf{not} overridden tasks}
            \For{$\tau_i\in\tau_\Omega$}
            \Comment{Assign \underline{overridden} tasks}
                \For{$\pi_i\in\pi_{sorted}$}
            \Comment{to \textbf{any} CPU core}
                    \If{$\Lambda_{\pi_i}=\emptyset$}
                        \State $\Lambda_{\pi_i}\gets\tau_i$
                        \State $\text{break}$
                    \EndIf
                \EndFor
            \EndFor
            \For{$\tau_i\in\tau'$}
            \Comment{Assign \textbf{not} overridden tasks}
                \For{$\pi_i\in\pi'$}
                \Comment{to CPU cores that \textbf{aren't} overheated}
                    \If{$\Lambda_{\pi_i}=\emptyset$}
                        \State $\Lambda_{\pi_i}\gets\tau_i$
                        \State $\text{break}$
                    \EndIf
                \EndFor
            \EndFor
            \State \Return $\{\Lambda_{\pi_i} \text{ where } \pi_i\in\pi \land \Lambda_{\pi_i}\neq\emptyset\}$
        \EndFunction
    \end{algorithmic}
\end{algorithm}

Algorithm~\ref{alg:needass} is used to check the conditions
of the tasks and CPU cores to see if an assignment is needed
either to ensure task completion, or to avoid overheating. It consists
of a set of five condition checks. The first condition checks to see if
any currently assigned tasks have completed, which triggers an
assignment. The second check checks to see if all of the tasks
in $\tau$ have completed. If so, no assignment is needed, since the
first condition would have already caused them to be unassigned. The
third condition checks to see if any tasks require an override
to ensure their completion by the end of the period. The fourth
check looks for any overheated cores that aren't currently idle
to trigger an assignment to idle them. Lastly, we check for
incomplete tasks in $\tau$ that are not currently assigned. If
such tasks exist, then if any cores are cool and idle, then an
assignment is triggered.

Algorithm~\ref{ch:vtftas:alg:assign} describes the task
assignment algorithm of VTF-TAS. Here, the candidate task
selection operations are integrated into the assignment
algorithm to enable the overriding of threshold restrictions
for tasks that must be executed continuously to meet their
deadline. First, the CPU cores are sorted according to their
temperature to create $\pi_{sorted}$. Also, a second sorted
set of CPU cores, $\pi'$ is created, excluding overheated
cores.  Next, the tasks are sorted according to their
remaining execution time to prioritize the execution of
tasks with higher remaining run time. $\tau_{\Omega}$ is a
set of the tasks that must be constantly executed in order
to meet their deadline, while $\tau'$ contains the tasks
without temperature overrides.  After these sets are
constructed, tasks in $\tau_\Omega$ are assigned to the
coolest available cores first to avoid unnecessary
overheating of the CPU cores. Then, any remaining tasks are
assigned to available cool CPU cores.

\begin{algorithm}[t]
    \caption{Pseudocode of Threshold Update Algorithm (Primary Concept)}
    \label{alg:threshup}
    \begin{algorithmic}[1]
        \Function{UPDATE\_THRESHOLD}{$\tau$, $T_H$, $p_{len}$, $t_{curr}$, $W_D$}
            \State \textbf{static} $C\gets0$; \textbf{static} $D\gets0$; \textbf{static} $D_{prev}\gets0$;
            \State $p_{done}\gets (t_{curr}\%p_{len})/p_{len}$
            \Comment{static variables keep their value between calls}
            \State $U_{\tau}\gets \frac{1}{n(\tau)}\sum_{i=1}^{n(\tau)}\frac{WCET(\tau_i)}{p_{len}}$
            \Comment{and are only instantiated to 0 on the first call}
            \State $U_{F}\gets -1*U_{\tau}*p_{done} + U_{\tau}$
            \State $U_{\sigma}\gets \frac{1}{n(\tau)}\sum_{i=1}^{n(\tau)}\frac{TimeLeft(\tau_i)}{p_{len}}$
            \State $H\gets (((U_{\sigma}/U_{F}) + (U_{\sigma}-U_{F}) )/2)-0.5$
            \State $H_{scaled}\gets H*(1-p_{done})$
            \Comment{Scales heuristic by remaining percent of period}
            \State $D\gets0$
            \If {$H_{scaled}>W_D$}
                \State $D\gets1$; $T_{Hnew}\gets T_H+(1/(C+1))$
            \ElsIf {$H_{scaled}<-1*W_D$}
                \State $D\gets-1$; $T_{Hnew}\gets T_H-(1/(C+1))$
            \EndIf

            \If {$D\neq0 \land (D_{prev}=0 \lor D=D_{prev})$}
                \State $C=C+1$
            \Else
                \State $C=0$
            \EndIf
            \State $D_{prev}\gets D$

            \State \Return $T_{Hnew}$
        \EndFunction
    \end{algorithmic}
\end{algorithm}

\begin{figure}[t]
    \centering
    \begin{subfigure}[t]{0.45\textwidth}
        \centering
        \includegraphics[width=1.0\textwidth]{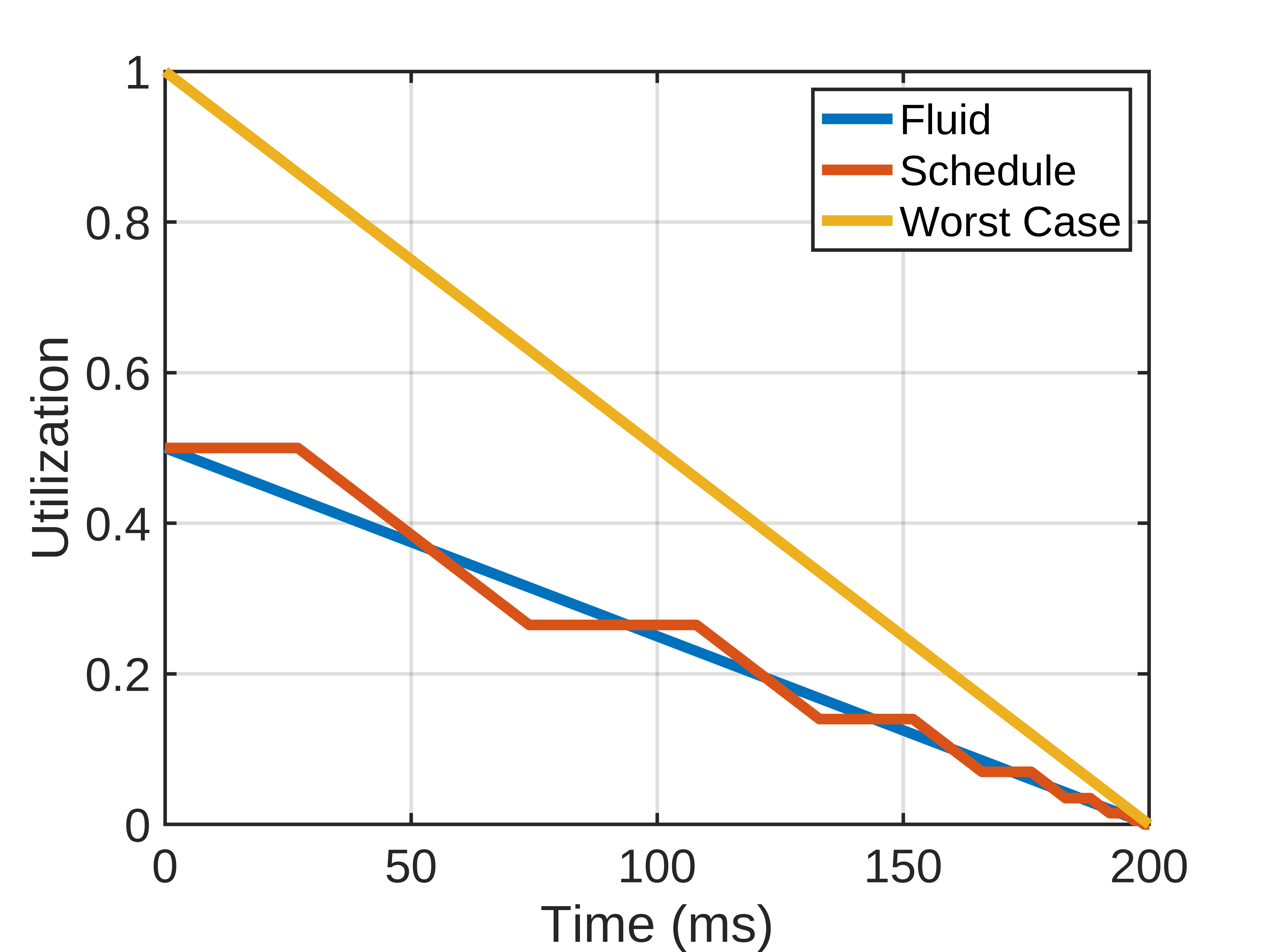}
        \subcaption{Utilization}
        \label{ch:vtftas:subfig:exutil}
    \end{subfigure}~
    \begin{subfigure}[t]{0.45\textwidth}
        \centering
        \includegraphics[width=1.0\textwidth]{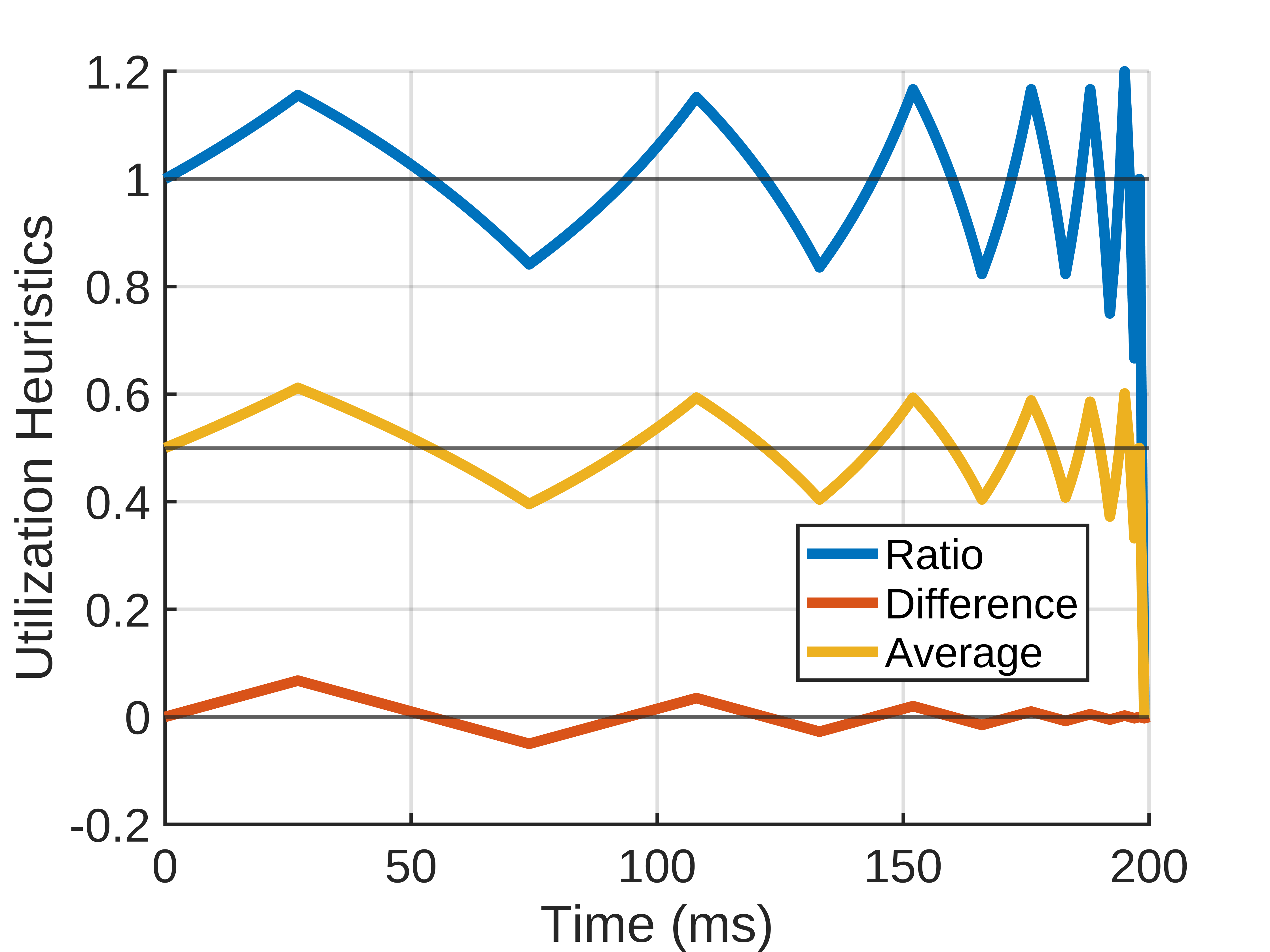}
        \subcaption{Heuristics}
        \label{ch:vtftas:subfig:exheur}
    \end{subfigure}
    \vspace{-0.5em}
    \caption{Examination of Utilization Behavior}
    \vspace{-1.5em}
    \label{ch:vtftas:fig:ex}
\end{figure}

The last algorithm used in the implementation of VTF-TAS is
Algorithm~\ref{alg:threshup}, which describes the process by
which $T_H$ is updated.  First, $C$, $D$, and $D_{prev}$ are
initialized to 0. These are static variables, so they keep
their value between calls to this algorithm. This is needed
to track the state of certain values throughout scheduling.
Then, the percentage of the current scheduling period that
has been completed is calculated to be used in other
calculations. $U_\tau$ stores the average of task
utilization of the period. Utilization is calculated as the
WCET of the task divided by the period length. The next
value, $U_{F}$ stores the remaining utilization of $\tau$ if
the tasks have been executing in a fluid manner.  $U_\sigma$
is calculated to be the average remaining utilization of the
task set. $\sigma$ is used for this notation, as this value
is dependent on the schedule.  Remaining utilization is
calculated as the remaining execution time of the task
divided by the period length.  $U_\sigma$ and $U_{F}$ are
then used to calculate $H$, the threshold update heuristic.
$H$ is calculated as the average of the difference and ratio
of $U_\sigma$ and $U_F$. We also subtract 0.5 from this
value so that when $U_\sigma=U_F$, $H=0$.  The average is
used to prevent the heuristic from being sensitive to the
ratio becoming too large or the difference becoming too
small at the end of the period. Averaging these two values
helps to keep $H$ stable throughout the period.  $H$ is also
scaled based on the remaining percent of the period, as the
ratio component can cause massive fluctuations in $H$ toward
the end of the period.

The scaled heuristic value is then compared to $W_D$, which
defines a ``dead zone'' around 0 for threshold updates. If
$H_{scaled}>W_D$, then the threshold is increased, otherwise
if $H_{scaled}<-1*W_D$, then the threshold is decreased.
Threshold updates are also dampened using $1/C+1$. $C$
stores the number of consecutive threshold updates in the
same direction. Thus, repeated updates to the threshold
cannot cause $T_H$ to become too large or too small. The
direction of the update is tracked with $D$, a positive
update results in $D=1$, while a negative update sets
$D=-1$. If $|H_{scaled}|<W_D$, then $D$ becomes 0 to denote
no update occurring. The final if statement in the algorithm
on line 14 checks to ensure that an update occurred, and that
it is either the first update in a given direction, or that
it is in the same direction as the previous update.

Figure~\ref{ch:vtftas:fig:ex} demonstrates the heuristic
behavior using a simulated execution schedule.
Figure~\ref{ch:vtftas:subfig:exutil} shows the utilization
of an example synthetic execution schedule (orange) versus
the fluid utilization (blue). When the schedule utilization
is not decreasing, this indicates that no tasks are
executing. Decrease of this value shows active task
execution.  Once the schedule utilization reaches 0, then
all tasks have completed their execution. The worst case
line indicates the point where the tasks would need to
constantly execute in order to complete before their
deadline. Should the schedule utilization exceed this, then
at least one tasks deadline will be breached.
Figure~\ref{ch:vtftas:subfig:exheur} shows the heuristics
that can be used for controlling the temperature threshold
during scheduling. We can see that the ratio of schedule and
fluid utilization is stable in the beginning of the period,
but suffers rapid fluctuations toward the end of the period.
Conversely, the difference of these values shows a loss of
sensitivity toward the end of the period. These effects are
due to the decreasing magnitude of the utilization as the
period continues. To rectify this issue, we can see that the
average of the ratio and difference gives a heuristic that
remains more stable throughout the entire period. This value
is used as $H$ in Algorithm~\ref{alg:threshup}.

\section{Evaluation Methodology}

For VTF-TAS, we improve the methodology from~\cite{dowling_2023}
to use a more realistic initial condition for temperature
prediction during TAS. By default, POD temperature
prediction begins with the CPU temperature being uniform at
$T_{amb}$, or 45$^\circ$C. While this is valid, it
represents a CPU that is entirely without power. This
scenario is not representative of a realistic TAS scenario,
where a CPU would be powered on, but idle, before beginning
schedule execution. Thus, we used the static power
dissipation of the CPU to perform POD prediction until the
output of the ODE solver stabilized. These ODE outputs, when
used to reconstruct the CPU temperature yield the steady
state temperature of the CPU when only static power is
applied.  This thermal state is representative of a CPU that
is idle, but powered on.

\begin{figure}[t]
    \centering
    \begin{subfigure}[t]{0.32\textwidth}
        \centering
        \includegraphics[width=1.0\textwidth]{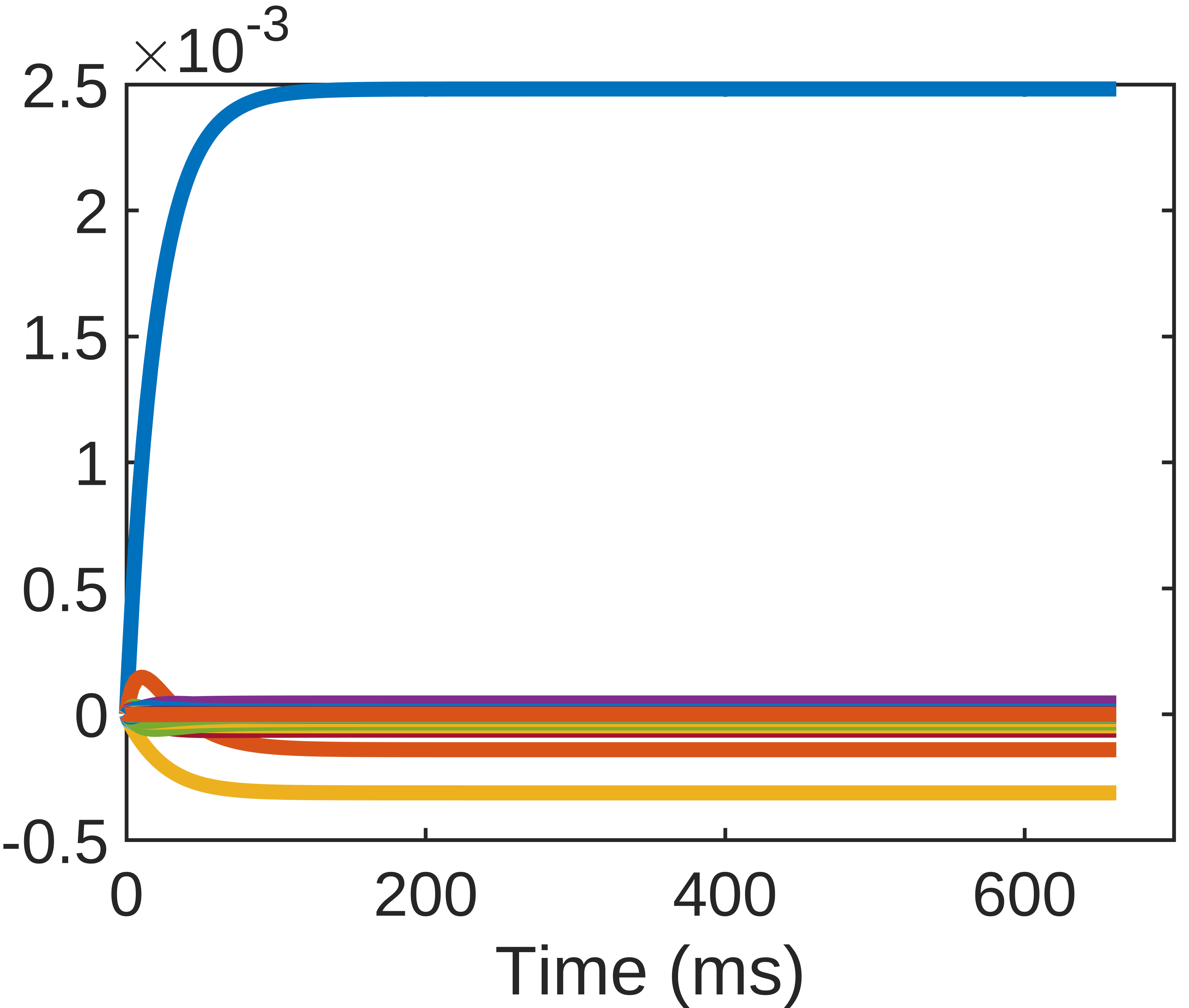}
        \subcaption{ODE Output}
        \label{ch:vtftas:subfig:ode}
    \end{subfigure}~
    \begin{subfigure}[t]{0.32\textwidth}
        \centering
        \includegraphics[width=1.0\textwidth]{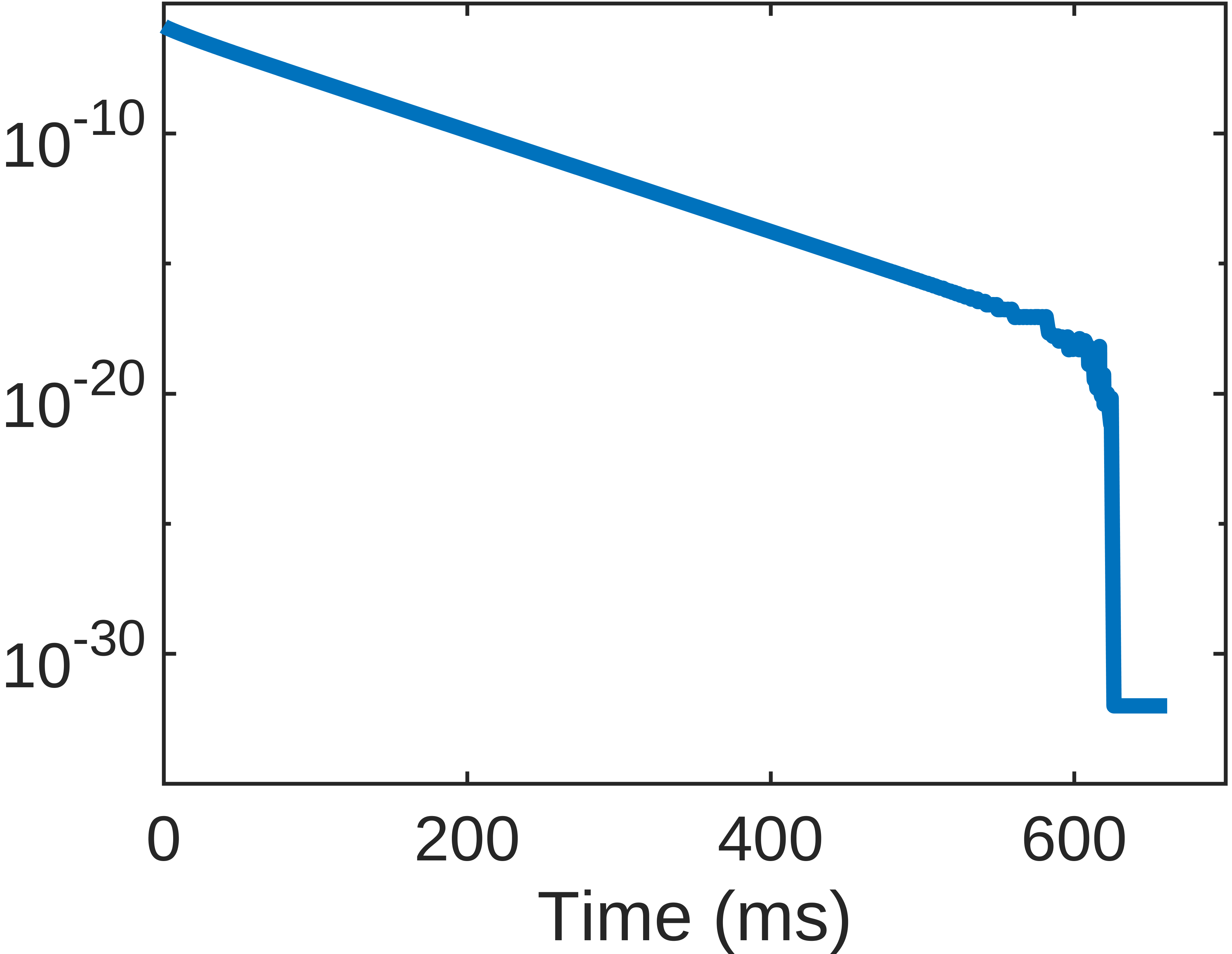}
        \subcaption{Max Absolute Gradient}
        \label{ch:vtftas:subfig:maxabs}
    \end{subfigure}~
    \begin{subfigure}[t]{0.32\textwidth}
        \centering
        \includegraphics[width=1.0\textwidth]{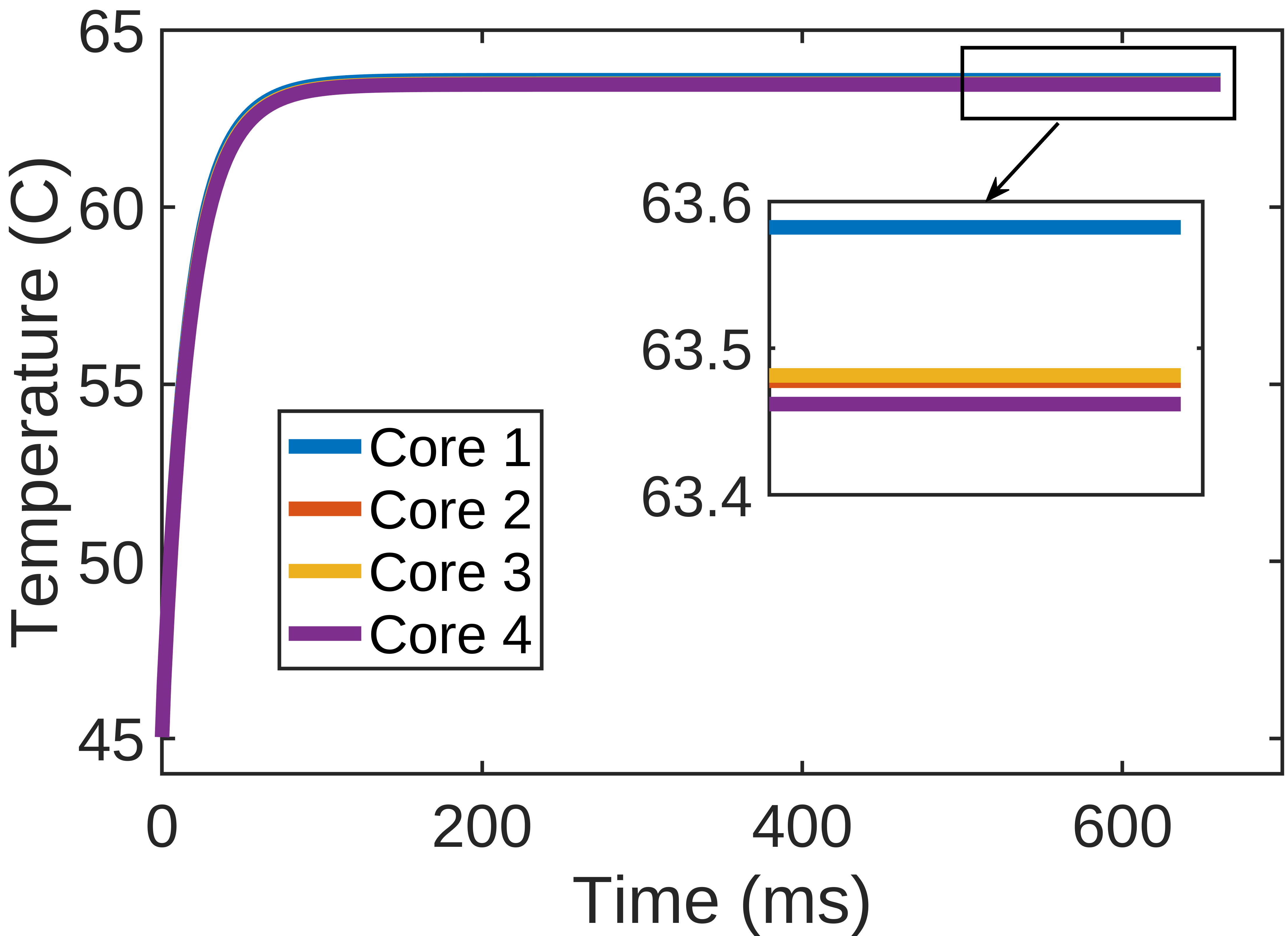}
        \subcaption{Reconstructed Temperature}
        \label{ch:vtftas:subfig:sstemp}
    \end{subfigure}
    \caption{Finding Idle Steady-State CPU Temperature}
    \label{ch:vtftas:fig:ss}
\end{figure}

Figure~\ref{ch:vtftas:fig:ss} shows this process.
Figure~\ref{ch:vtftas:subfig:ode} shows the calculated ODE
coefficients. They appear to stabilize very quickly, but
Figure~\ref{ch:vtftas:subfig:maxabs} shows the maximum
absolute value of the first derivative of the coefficients.
This shows that it takes much longer for their values to
stop changing. Around 600ms, we can see that it drops to a
value of $10^{-32}$ and does not change. In the computed
data, we replaced 0 values with this value to enable
logarithmic scale plotting of the data.  Thus, the ODE
output completely stops changing at this point.
Figure~\ref{ch:vtftas:subfig:sstemp} shows the reconstructed
temperature of the CPU cores in this case. We can see that
they stabilize to nearly the same value. Since our CPU
floorplan is symmetric, this is expected. However, the
values are not perfectly equal due to numerical error during
FEM and POD training.

\begin{figure}[ht]
    \centering
    \includegraphics[width=0.90\textwidth]{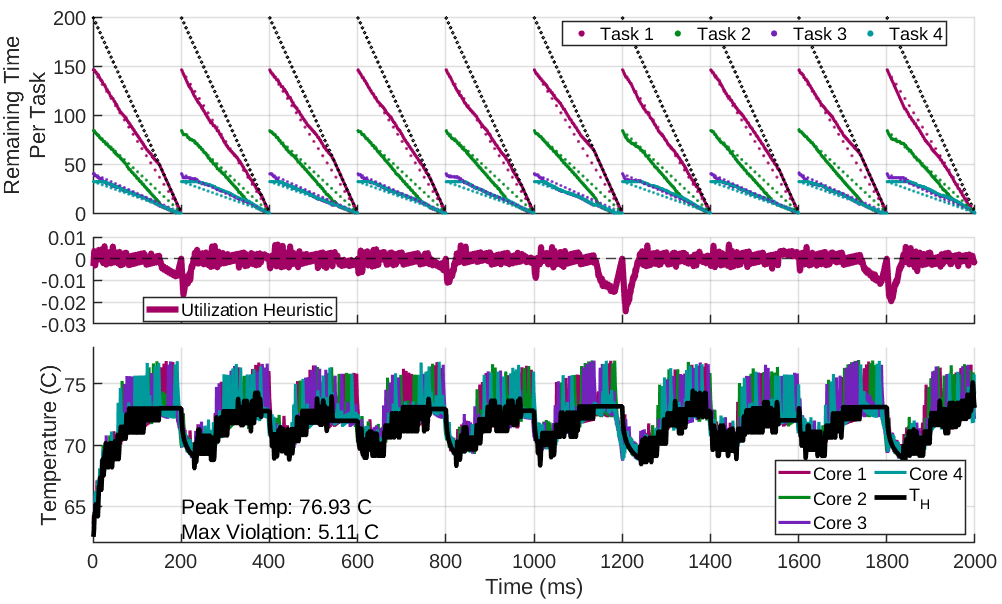}
    \caption{VTF-TAS with $W_D=0.0$}
    \label{ch:vtftas:fig:wdo}
\end{figure}

\section{Evaluating VTF-TAS}
\label{ch:vtftas:sec:eval}

To evaluate the VTF-TAS algorithm's scheduling performance,
we have constructed schedules under a number of conditions
to observe the predicted thermal behavior. This evaluation
gives us an understanding of the behavior of VTF-TAS and due
to the accuracy of the POD thermal model, we can assume that
the actual CPU temperature when calculated with FEM will
differ minimally from the temperature predicted during
scheduling.

The first experiment performed is with $W_D=0$. The results
of this test are shown in Figure~\ref{ch:vtftas:fig:wdo}.
The topmost subplot shows the remaining execution time for
each task throughout the scheduling period. The dotted lines
represent the remaining execution times of the task if it
were executing according to a purely fluid schedule. The
black dotted line represents the remaining time in the
period. Should a task's remaining run time meet this line,
then that task is given an override, which disables
threshold updates, and allows that task to execute on any
core, even overheated ones, to ensure its completion before
its deadline. We tend to see overrides occurring towards the
end of each period due to task 1 executing too slowly,
triggering the override mechanism.  The middle plot shows
the scheduling heuristic. In general, if tasks are executing
more quickly than their fluid execution rate, then this
heuristic is below 0, while slower execution brings the
heuristic above 0.  Since in this experiment, $W_D=0$, when
the heuristic is negative, $T_H$ decreases, and vice versa.
The last plot shows the predicted CPU temperature during
scheduling for each core along with $T_H$ being shown as the
solid black line. $T_H$ tends to be regularly updated in
this scenario, but remains fairly stable throughout
scheduling.

\begin{figure}
    \centering
    \includegraphics[width=0.9\textwidth]{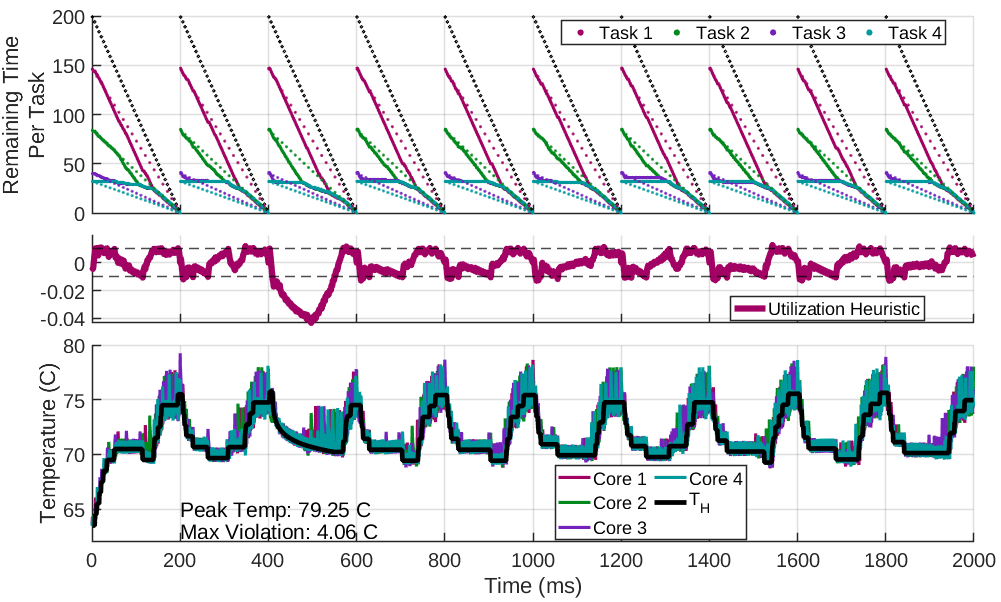}
    \vspace{-0.5em}
    \caption{VTF-TAS with $W_D=0.01$}
    \vspace{-1em}
    \label{ch:vtftas:fig:wdo1}
\end{figure}

Figure~\ref{ch:vtftas:fig:wdo1} shows the scheduling
behavior with $W_D=0.01$. Here, the heuristic is allowed to
vary around 0 before triggering an update to $T_H$, which
leads to a generally more stable $T_H$ value. However, we
can see that this leads to high variance in $T_H$ since
toward the end of the period, many fast updates to increase
$T_H$ occur, causing a spike, leading to a worse overall
peak temperature, and higher variance in the core
temperatures.

\begin{table}
    \centering
    \caption{Effects of $W_D$ on VTF-TAS Performance}
    \vspace{-0.5em}
    \label{ch:vtftas:tab:wd}
    \begin{tabular}{|c|c|c|}
        \hline
        $W_D$   & Peak Temp ($^\circ$C)& Max Violation ($^\circ$C) \\ \hline\hline
        0.0     & 76.93 & 5.11 \\ \hline
        0.001   & 77.07 & 5.13 \\ \hline
        0.01    & 79.25 & 4.06 \\ \hline
        0.1     & 107.43& 36.93 \\ \hline
    \end{tabular}
    \vspace{-0.5em}
\end{table}

Table~\ref{ch:vtftas:tab:wd} shows the peak CPU temperature
and maximum threshold violations for varying values of
$W_D$. From this table, we can see that maximum performance
is achieved when $W_D=0$, meaning that while this mechanism
can help prevent unnecessary changes to $T_H$, it does have
a slightly negative impact on overall thermal performance of
the constructed schedule. From this evaluation, we can see
that when $W_D=0$, this algorithm's peak temperature is
lower than POD-TAS'~\cite{dowling_2023} peak temperature by
$\approx1.5^\circ$C. Plots similar to
Figures~\ref{ch:vtftas:fig:wdo} and~\ref{ch:vtftas:fig:wdo1}
for $W_D=0.1$ and $W_D=0.001$ are shown
in~\ref{sec:appendix}.  We do not perform the final FEM
evaluation of the schedule temperature in this work due to
the high accuracy demonstrated by the POD thermal model used
during scheduling.

\section{Conclusion}
\label{sec:conc}

Using the methodology and core concepts put forth with the
POD-TAS algorithm~\cite{dowling_2023}, we designed and
evaluated a new algorithm for TAS, VTF-TAS. This algorithm
utilizes concepts from fluid scheduling to enable the
adaptive management of the temperature threshold during task
scheduling. This capability removes the need for the search
process used for POD-TAS~\cite{dowling_2023}, removing a high computational
overhead for obtaining optimal thresholds to use during
scheduling. Through our evaluation, we demonstrate that
VTF-TAS is capable of achieving a lower peak temperature
than the POD-TAS algorithm while not violating task
deadlines.

\bibliographystyle{plain}
\bibliography{vtftas}

\begin{thebibliography}{1}

\bibitem{dowling_2023}
Anthony Dowling, Lin Jiang, Ming-Cheng Cheng, and Yu~Liu.
\newblock Regulating cpu temperature with thermal-aware scheduling using a
  reduced order learning thermal model.
\newblock {\em arXiv preprint arXiv:2310.00854}, 2023.

\bibitem{combs_2020}
Anthony Dowling, Frank Swiatowicz, Yu~Liu, Alexander~John Tolnai, and
  Fabian~Herbert Engel.
\newblock Combs: First open-source based benchmark suite for multi-physics
  simulation relevant hpc research.
\newblock In {\em Algorithms and Architectures for Parallel Processing: 20th
  International Conference, ICA3PP 2020, New York City, NY, USA, October 2--4,
  2020, Proceedings, Part I 20}, pages 3--14. Springer, 2020.

\bibitem{fluid_2019}
Muhammad~Naeem Shehzad, Qaisar Bashir, Ghufran Ahmad, Adeel Anjum,
  Muhammad~Naeem Awais, Umar Manzoor, Zeeshan~Azmat Shaikh, Muhammad~A
  Balubaid, and Tanzila Saba.
\newblock Thermal-aware resource allocation in earliest deadline first using
  fluid scheduling.
\newblock {\em International Journal of Distributed Sensor Networks},
  15(3):1550147719834417, 2019.

\end{thebibliography}

\appendix
\section{VTF-TAS with Other Values of $W_D$}
\label{sec:appendix}

\begin{figure}[h!]
    \centering
    \includegraphics[width=0.9\textwidth]{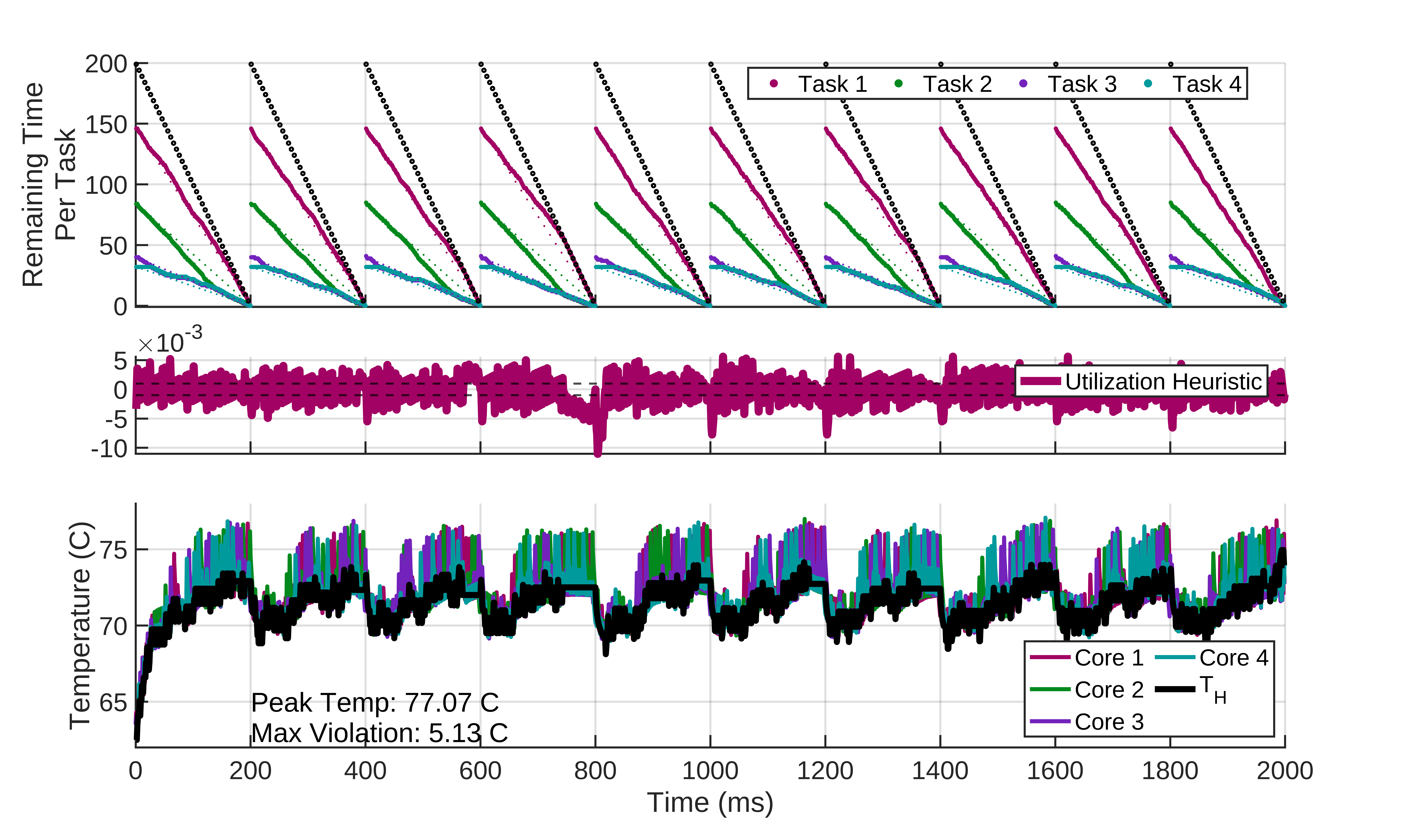}
    \caption{VTF-TAS with $W_D=0.001$}
    \label{ch:vtftas:fig:wdoo1}
\end{figure}

\begin{figure}[h!]
    \centering
    \includegraphics[width=0.9\textwidth]{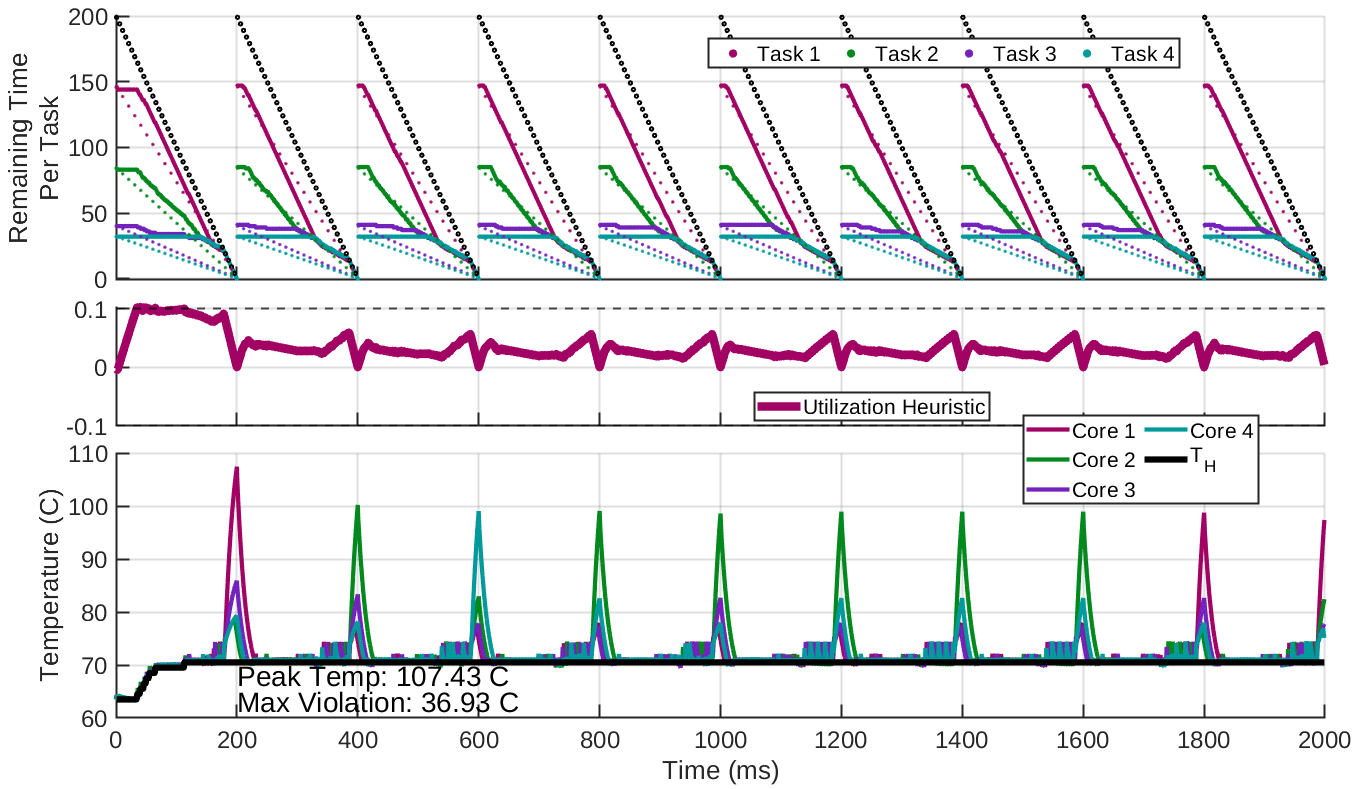}
    \caption{VTF-TAS with $W_D=0.1$}
    \label{ch:vtftas:fig:wd1}
\end{figure}

\end{document}